\documentclass[12pt]{article}
\newcommand{\tps}{\tilde{\psi}}
\newcommand{\tH}{\tilde{H}}

\newcommand{\tc}{\tilde{\chi}}
\newcommand{\suc}{supersymmetric\ }

\newcommand{\del}{\partial}
\newcommand{\be}{\begin{equation}}
\newcommand{\ee}{\end{equation}}
\newcommand{\bea}{\begin{eqnarray}}
\newcommand{\eea}{\end{eqnarray}}
\newcommand{\ba}{\begin{array}}
\newcommand{\ea}{\end{array}}
\newcommand{\beas}{\begin{eqnarray*}}
\newcommand{\eeas}{\end{eqnarray*}}
\newcommand{\bes}{\begin{equation*}}
\newcommand{\ees}{\end{equation*}}

\newcommand{\nn}{\nonumber}
\newcommand{\lf}{\left}
\newcommand{\ri}{\right}
\newcommand{\f}{\frac}

\def\cL{{\cal L}}
\def\cD{{\cal D}}

\def\tr           {\mbox{\rm tr}\,}
\def\det           {\mbox{\rm det}\,}
\def\ha           {\mbox{$\frac{1}{2}$}}
\def\stw           {\mbox{$\sqrt{2}$}}
\def\tht           {\mbox{$\frac{3}{2}$}}
\def\qu           {\mbox{$\frac{1}{4}$}}
\def\al           {\alpha}

\def\bet           {\beta}
\def\beb           {{\bar \beta}}
\def\ch           {\chi}
\def\de           {\delta}
\def\ep           {\epsilon}
\def\vep           {\varepsilon}
\def\et           {\eta}
\def\ga           {\gamma}

\def\la           {\lambda}

\def\om           {\omega}
\def\ph           {\phi}
\def\ps           {\psi}
\def\rh           {\rho}
\def\si           {\sigma}
\def\ta           {\tau}

\def\ze           {\zeta}
\def\pl           {\partial}

\def\ran          {\rangle}
\def\lan          {\langle}
\def\na           {\nabla}
\def\cb           {{\bar c}}
\def\wb           {{\bar w}}
\def\zb           {{\bar z}}
\def\mb           {{\bar{m}}}

\def\gr          {{\sqrt g_1}}
\def\gs          {{\sqrt g_2}}
\def\gb          {{\bar g}}

\def\bon{{\bf 1}}
\def\btw{{\bf 2}}
\def\bth{{\bf 3}}

\def\bsi{{\bf 6}}

\begin{document}

\title{ $N=4$ SYM on $\Sigma \times S^2$ and its  
Topological Reduction}
\author{ A Imaanpur \thanks{Email: aimaanpu@physics.adelaide.edu.au}\\
{\small {\em Department of Physics and Mathematical Physics}} \\
{\small {\em University of Adelaide, Adelaide, SA 5005, Australia}}
}
\maketitle
 
\begin{abstract}
We consider the twisted  $N=4$ SYM on $\Sigma \times S^2$. In the 
limit that $S^2$ shrinks to zero size the four dimensional theory reduces 
to a two dimensional SYM theory. We compute the correlation functions of a 
set of BRST cohomology classes in the  
reduced theory perturbed by mass. 

\vspace{5 mm}

\end{abstract}

\section{Introduction} 

Topological field theories \cite{WTFT}  have proven to be a   useful tool in 
the investigation of the 

nonperturbative characteristics of  supersymmetric gauge theories. There is 
 an interplay between certain \suc gauge theories and their corresponding topological 
versions:  
 one 

can use topological results on smooth manifolds to learn about the underlying 
physical theory; conversely, one may   use the physical arguments to gain new 
insight into the 
topological structure of the manifold on which fields are defined \cite{WS}.

As an example of the first -- i.e.\ , using the results 

on the mathematical side to learn about  physics --  consider the $N=4$ SYM theory. 
This theory has been conjectured to have an exact $SL(2, {\bf Z})$ duality \cite{MO}.
 Since this duality relates  the weak and strong coupling behaviour of 
the theory, to test the conjecture  one needs quantities such as the partition 
function to be computed nonperturbatively. This is 

a formidable task and one actually does not know how to proceed in this direction. 
This is where topological field theory comes to provide an alternative approach 
to the problem. Instead of the physical theory, one considers 
the corresponding topological field theory 

obtained by a procedure called twisting. The basic characteristics of the theory, 
such as 

$SL(2, {\bf Z})$ invariance, remain intact under twisting, so one hopes to see the 
realization of this  symmetry in the twisted model. In  \cite{VW} it has been shown, 
 using the known facts about the 
structure of the moduli space of instantons and the associated Euler characteristic,     
the partition function of $N=4$ twisted theory on some specific manifolds can be 
computed. So in this way  
 it has become possible  
to make some exact and nonperturbative statements about the theory and its self-duality 
properties. 

In this article, we will study the $N=2$ reduction of the above theory obtained 
 by mass perturbation  for the hypermultiplet. This theory  
is still believed 
to be $S$-dual \cite{SW}. We will compute the correlation functions of 
a set of specific operators using 
a method of calculation  similar to that of \cite{WRE}.  

Twisted $N=2$ and $N=4$ SYM theories on product manifolds $\Sigma \times C$, where $\Sigma $ 
and 
$C$ are both Riemann surfaces, have been considered in \cite{BJSV}. There it was shown 
that, in the limit $C$ shrinks to zero, the four dimensional theory generically 
reduces to an 
effective two dimensional sigma model. However, when $C$ is a Riemann sphere -- as 
is the case of interest in the
present paper -- things are 
a bit different. The   
dimension of the self-dual harmonic 2-forms, $b_2^+$, is one in this case. Hence 
 the  connection is reducible.  
It follows then that the path integral may get contribution from the so called $u$-plane 
\cite{MW, NEK}. Moreover, when $b_2^+=1$, there is a wall in the space of one parameter metrics.  
On crossing this wall the partition function may change its value. 

Here  we will compute the path integral in a chamber where $S^2$ shrinks to zero. 
We  consider $SO(3)$  bundles such that the restricted bundle over  
$S^2$ is trivial.  Bundles which restrict nontrivially on $S^2$ give 
zero contribution 
to the path integral. This is so because in the limit that $S^2$ shrinks to zero size, 
the path 
integral localizes on the moduli space of flat connections in the $S^2$ direction. 
However, it can be shown that for a flat bundle over $S^2$, transition functions are 
trivial and the bundle must be trivial. 
Therefore nontrivial $SO(3)$ bundles on $S^2$     
do not admit flat connections. 

The organization of this paper is as follows: In section 2 we consider the twisted 
$N=4$ Lagrangian on $\Sigma\times S^2$. In the limit where $S^2$ shrinks it is 
shown how  the four dimensional theory reduces to an effective two dimensional 
theory. The fixed point equations imply, in the case of a nontrivial $SO(3)$ 
bundle over $\Sigma$, that the partition function of this reduced theory is in fact  
the Euler characteristic of the moduli space of flat connections on $\Sigma$. 
A mass perturbation  makes the path integral calculation more tractable -- particularly  
for the limiting two-dimensional theory.  In section 3, we show how this comes about. 
Perturbing 
by the mass allows most fields to be integrated out,  
and reduces the path integral to a finite 
dimensional integral which can be easily performed. In section 4 we 
discuss the result. Although we have not yet managed to give an 
explicit check of $S$-duality, we have isolated the problems involved and hope 
to return to this in later work.

\section{Twisted $N=4$ on $\Sigma \times S^2$ and its reduction} 

The key point in twisting \cite{WTFT} is to redefine the global space-time symmetry 
such that at least one component of the supercharge becomes scalar under the 
new defined space-time symmetry. This procedure crucially depends on the 
existence of a suitable global R-symmetry. $N=4$ SYM theory in four dimensions has 
a global $SU(4)$ symmetry such that the supercharges transform under the ${\bf 4}$ of 
this symmetry. First one needs to see how this representation transforms under the 
space-time symmetry group $SU(2)_L\times SU(2)_R$. There are  \cite{VW} three 
possibilities for the  
decomposition which give rise to singlets under the twisting: (i)$\ (\btw , \bon)\oplus 
(\bon , \btw);$ (ii)$\ (\bon , \btw)\oplus (\bon , \btw);$ (iii)$\ (\bon , \btw)\oplus 
(\bon , \bon) \oplus (\bon , \bon)$. As in \cite{VW}, we will consider the case 
(ii) where, after twisting,  two components of the supercharges turn out to be singlets 
and therefore square to zero. The scalar fields of the physical theory,
 which transform under the $\bsi$ of $SU(4)$, now 
transform   
under the new rotation group,  $SU(2)_L\times SU(2)'_R$, as $3 (\bon , \bon) 
\oplus (\bon , \bth)$, three singlets and one self-dual 2-form.

Having determined how the new fields transform under the new 
 symmetry 
group, what remains is  to rewrite the Lagrangian in terms of these new 
fields on flat ${\bf R^4}$. This Lagrangian  can then be defined on an 
arbitrary smooth four manifold while preserving those two BRST like symmetries.

Let us start our discussion with the twisted  $N=4$ Lagrangian\footnote{The Lagrangian 
that we use is actually  
different from the one constructed in \cite{L} by a BRST exact term 
$ - \frac{i}{4}\de(\et[\ph ,\la])$.} in 4 dimensions \cite{VW,L}, 
\bea
\cL & = & \frac{1}{e^2}\tr \left\{ - D_\mu \la D^\mu \ph +\ha \tH ^\mu (\tH 
_\mu -2\stw D_\mu C +4\stw D^\nu B_{\nu \mu}) \right. \nn \\
 & + & \ha H^{\mu \nu}(H_{\mu \nu}
-2 F_{\mu \nu}^+ -4i[B_{\mu \rh},B^\rh _{\ \nu}] -4i[B_{\mu \nu},C]) \nn \\
 & + & 4 \ps_\mu D_\nu \ch^{\mu \nu} + 4\tc_\mu D_\nu \tps^{\mu \nu} + 
\tc_\mu D^\mu\ze - \ps_\mu D^\mu\et \nn \\
 & + & i\stw \tps^{\mu \nu}[\tps_{\mu \nu},\la ] -i\stw \ch^{\mu \nu}
[\ch_{\mu \nu},\ph] + i 2\stw \tps^{\mu \nu}[\ch_{\mu \nu},C]+i 4\stw 
\tps^{\mu \nu}[\ch_{\mu \rh}, B_\nu ^{\ \rh}] \nn \\
 & - & i\stw \ch_{\mu \nu}
[\ze ,B^{\mu \nu}] -i\stw \tps_{\mu \nu}[\et ,B^{\mu \nu}]+ i4\stw 
\ps_\mu [\tc_\nu ,B^{\mu \nu}]
- i\stw \tc_\nu [\tc^\nu, \ph ]   \nn \\
 & + & i\stw \ps_\mu [\ps^\mu, \la ] - i2\stw \ps_\mu [\tc^\mu ,C ]+
\frac{i}{2\stw }\ze [\ze, \la] \nn \\
 & - &\left. \frac{i}{\stw } \ze [\et , C]+ 2[\ph ,B^{\mu \nu}]
[\la , B_{\mu \nu}] + 2 [\ph ,C][\la ,C] \right\}.        
\eea
As mentioned, the action is invariant under two BRST transformations. However, 
for us it is 
enough to consider one of them, which reads \cite{L}
\bea \begin{array}{lll}
 &\de A_\mu  =  -2\ps_\mu &  \de \ze =4i[C,\ph ] \nn \\
 &\de \ps_\mu =  -\stw D_\mu \ph  &  \de \la = \stw \et \nn \\
&\de \ph =0&  \de \et = 2i [\la ,\ph ] \nn \\
 &\de B_{\mu \nu} = 
 \stw \tps_{\mu \nu} &  \de \tc_\mu = \tH_\mu \nn \\
& \de \tps_{\mu \nu} = 
 2i[B_{\mu \nu},\ph ] &  \de \tH_\mu =2\stw i 
[\tc_\mu ,\ph ] \nn \\
 &\de C =  \frac{1}{\stw}\ze & \de \ch_{\mu \nu}= H_{\mu \nu} \nn \\
 &  & \de H_{\mu \nu} 
= 2\stw i [\ch_{\mu \nu}, \ph ] . 
\end{array}
\eea
In this article we choose $\ph$ and $\la$ to be two independent real 
scalars. This will render the Lagrangian to be hermitian and allow us 
to treat $\ph$ and $\la$ independently. The generators of the $SU(2)$ group 
are chosen to be hermitian $T^a =\f{1}{\stw}\si^a$ with $\tr (T^aT^b)=\de ^{ab}$. 
 
The theory enjoys an exact $U(1)$ ghost symmetry under which $\ps_\mu , 
\tps_{\mu \nu} , \ze$ have charge 1, $\ch_{\mu \nu}, \et , \tc_\mu$ charge 
$-1$,  while $\ph$ and $\la$ have charges 2 and $-2$ respectively. All other 
fields have ghost number zero.

Take the underlying manifold to be $\Sigma \times S^2 $. 
Let us denote the indices on $\Sigma$ by 
$i,j,\cdots $ and those on $S^2$ by $a,b,\cdots $. We define
\bea
F_{ij}=\f{1}{\sqrt g_1}\ep_{ij} f  & & \ch_{ij}=\f{1}{\gr}\ep_{ij} \ch  \nn \\
 B_{ij}=\f{1}{2\gr} \ep_{ij} b & & \tps_{ij}=\f{1}{2\gr} \ep_{ij} \tps\  , \label{defi1}
\eea
and the same for indices on $S^2$
\be
B_{ab}=\f{1}{2\gs} \ep_{ab} b' \ ,\ \  \ch_{ab}=\f{1}{\gs}\ep_{ab} \ch '  \ ,
\ \   \tps_{ab}=\f{1}{2\gs} \ep_{ab} \tps ' . 
\label{defi2}
\ee
Here $g_1$ and $g_2$ denote the determinant of the metric on $\Sigma$ and $S^2$ respectively.

The fields $H_{\mu\nu}$, $B_{\mu\nu} ,\ \ch_{\mu\nu}$ and $\tps_{\mu\nu}$ are all  self-dual. 
Note that
\bea
B_{ij} = *B_{ij} \Rightarrow \f{1}{2\gr} \ep_{ij} b & = & \frac{1}{2\sqrt g} 
\ep_{ij}^{\ \ ab} B_{ab}=
\frac{1}{2\sqrt g} \ep_{ij}\ep^{ab} (\f{1}{2\gs} \ep_{ab}b') \nn \\
 & = & \frac{1}{4\sqrt {g g_2}} 2 g_2\ \ep_{ij} b'= \f{1}{2\gr} \ep_{ij} b'\ ,
\eea
where we have used that 
$
\ep^{ab}\ep_{ab}=\ep^{ab}\ep^{a'b'}g_{aa'}g_{bb'}=2g_2 $ and $
g^{aa'}g^{bb'}\ep_{a'b'}=\ep^{ab} $.
Also we chose  $\ep^{12}=1$ and so $\ep_{12}=g_2$; thus, for example, we have 
$B^{ab}=\f{1}{2\gs} \ep^{ab} b'$.
Hence we conclude that 
\[
b=b'\ ,\ \ \ch =\ch '\ ,\ \ \tps =\tps '.
\]

In \cite{BJSV} it was shown that upon  shrinking the metric on $\Sigma$, one gets 
an effective 2-dimensional sigma model governing the maps from $S^2$ to $\cal M$, where 
$\cal M$ is the moduli space of solutions to the Hitchin's equations. Although the 
twisted theory is supposed to be topological,  
since the space of self-dual harmonic forms in this case is one-dimensional one 
may not get the same effective theory if one instead shrinks $S^2$. In that case 
 we will see  that the effective  theory which emerges    
is  a 2-dimensional twisted SYM theory, as conjectured in \cite{BJSV}. 

Thereto, we now scale the metric on 
$S^2$ by a factor of $\ep$. Notice that   
the definitions (\ref{defi1}) and (\ref{defi2})  are consistent with this scaling,  
 since both sides of the self-duality  
constraints scale with the same power of $\ep$.
%
%
%

\noindent
After  integrating out the auxiliary fields,  the bosonic part of the 
Lagrangian reads
\be
\cL_{\rm B}  = \frac{1}{e^2}\tr \left\{ - D_\mu \la D^\mu \ph 
-( D_\mu C -2 D^\nu B_{\nu \mu})^2 
-\ha( F_{\mu \nu}^+ +2i[B_{\mu \rh},B^\rh _{\ \nu}] +2i[B_{\mu \nu},C])^2 \ri\},
\label{bosonic}
\ee
where $F^+=\ha (F+*F)$ and  $*$ is the Hodge duality operation. Thus we can write
\[
-\ha\int {\sqrt g} F_{\mu \nu}^+F^{\mu \nu +} =-\qu\int 
{\sqrt g}F_{\mu \nu}F^{\mu \nu}-\qu \int {\sqrt g}\ (*F)_{\mu \nu}F^{ \mu \nu}.
\]
The last term is the instanton number and is metric independent. Using this,  
and the fact that $B_{\mu \nu}$ is self-dual, we write the last term in 
(\ref{bosonic}) as 
\bea
&-& \qu F_{\mu \nu}F^{\mu \nu}-2iF^{\mu \nu}([B_{\mu \rh},B^\rh _{\ \nu}] 
+[B_{\mu \nu},C])
+2 ([B_{\mu \rh},B^\rh _{\ \nu}] +[B_{\mu \nu},C])^2 -\qu
 (*F)_{\mu \nu}F^{ \mu \nu}  \nn \\
&=&
-\qu\lf\{ F_{ij}F^{ij}+8i F^{ij}([B_{ij} , C] + [B_{ia} , B^a_{\ j}]) 
-8([B_{ij} , C] + [B_{ia} , B^a_{\ j}])^2 + (*F)_{ij}F^{ij} \ri.
\nn \\
&+& \lf. F_{ab}F^{ab}+8i F^{ab}([B_{ab} , C] +[B_{ai} , B^i_{\ b}]) 
-8 ([B_{ab} , C] +[B_{ai} , B^i_{\ b}])^2 + (*F)_{ab}F^{ab} \ri\}
\nn \\
&-&
(F_{ai}^+ +2i[B_{aj},B^j _{\ i}] +2i[B_{ab},B^b_{\ i}] +2i[B_{ai},C])^2 
\nn \\
&=& 
-\qu(F_{ij} +4i[B_{ij} , C])^2 -\qu(F_{ab} +4i [B_{ai} , B^i_{\ b}])^2 
-2iF^{ij}[B_{ia} , B^a_{\ j}] -2i F^{ab}[B_{ab} , C] 
\nn \\
&+&
4[B^{ij} , C][B_{ia} , B^a_{\ j}] +4[B^{ab} , C][B_{ai} , B^i_{\ b}]
-\qu (*F)_{ij}F^{ij} -\qu (*F)_{ab}F^{ab}) 
\nn \\
&-&
(F_{ai}^+ +2i[B_{aj},B^j _{\ i}] +2i[B_{ab},B^b_{\ i}] +2i[B_{ai},C])^2 . 
\nn
\eea
In the last equality we noted that for a  
self-dual antisymmetric tensor we have $S_{ij}^2 =S_{ab}^2$. In particular 
\bea
&& [B_{ab} , C]^2 =[B_{ij} , C]^2 \nn \\
&& \tr ([B_{ai} , B^i_{\ b}][B^{aj} , B_j^{\ b}]) 
= \tr ([B_{ia} , B^a_{\ j}][B^{ib} , B_b^{\ j}]). \nn
\eea

After scaling the metric, then, the Lagrangian splits to three parts; 
\[
\cL =\cL_{1} +\cL_0 +\cL_{-1},
\]
where $\cL_n$ scales as $\ep^n$. Specifically, 
\bea
\cL_1 & = & \frac{\ep}{e^2} \tr \left\{ - D_i \la D^i \ph  - D_i C D^i C 
- D_i b D^i b +\f{2}{\gr}\ep^{ij} D_i bD_j C -\f{1}{2}(f+2i[b,C])^2 \right. \nn \\
 & + &\f{4}{\gr} \ep^{ij}\ps_i D_j \ch + \f{2}{\gr}\ep^{ij} \tc_i D_j \tps + 
\tc_i D^i\ze - \ps_i D^i\et \nn \\
 & + & i{\stw} \tps [\tps ,\la ] -i4\stw \ch
[\ch ,\ph] + i 4\stw \tps [\ch ,C]  -2i\stw \ch
[\ze ,b] \nn \\
 & - & i{\stw} \tps [\et ,b] + i \f{2\stw}{\gr} \ep^{ij} \ps_i 
[\tc_j ,b] -i\stw \tc_i [\tc^i , \ph ] 
  + i\stw \ps_i [\ps^i , \la ] \nn \\ 
& - &   i2\stw \ps_i [\tc^i ,C ] + \frac{i}{2\stw }\ze 
[\ze , \la]  -  \frac{i}{\stw } \ze [\et , C] \nn \\
 & + & \left. 2[\ph ,b][\la , b] + 2 [\ph ,C][\la ,C]  \right\},
\eea
\bea
\cL_0  & = & \frac{1}{e^2}\tr \left \{ - D_a \la D^a \ph -( D_a C  
+\f{1}{\gs}\ep_{ab}D^b b -2D^i B_{ia})^2 +4(D^aB_{ai})(D^iC-2D^jB_j^{\ i}) \right. \nn \\
 & - & (F_{ai}^+ +2i[B_{aj},B^j _{\ i}] +2i[B_{ab},B^b_{\ i}] +2i[B_{ai},C])^2 
-\qu (*F)_{ij}F^{ij} -\qu (*F)_{ab}F^{ab} 
\nn \\
&-& 2i(F_{ij}+2i[B_{ij},C])[B^{ia},B_a^{\ j}] 
-2i (F_{ab} +2i[B_{ai} , B^i_{\ b}])[B^{ab} , C]
\nn \\
 & + & 4 \ps_{[i} D_{a]} \ch^{ia} + 4\tc_{[i} D_{a]} \tps^{ia} + 
\tc_a D^a\ze - \ps_a D^a\et + 4 \ps_{a} D_{b} \ch^{ab}
 + 4\tc_{a} D_{b} \tps^{ab}  \nn \\
 & + & 2 i\stw \tps^{ai}[\tps_{ai},\la ] -2i\stw \ch^{ai}
[\ch_{ai},\ph] + i 4\stw \tps^{ai}[\ch_{ai},C]+i 4\stw \tps^{ab}
[\ch_{ai}, B_b ^{\ i}] \nn \\
 & + &  i4\stw \tps^{ia}
[\ch_{ib}, B_a ^{\ b}]+ i 4\stw \tps^{ai}[\ch_{ab}, B_i ^{\ b}]+i 
4\stw \tps^{ai}[\ch_{aj}, B_i ^{\ j}] + i4\stw \tps^{ia}
[\ch_{ij}, B_a ^{\ j}] \nn \\
 & + & i 4\stw \tps^{ij}[\ch_{ia}, B_j ^{\ a}]- 2i\stw \ch_{ai}
[\ze ,B^{ai}] -2i\stw \tps_{ai}[\et ,B^{ai}]  
+i4\stw \ps_a [\tc_b ,B^{ab}]\nn \\ 
 & + & i4\stw \ps_a [\tc_i ,B^{ai}] + i4\stw \ps_i [\tc_a ,B^{ia}]
- i\stw \tc_a [\tc^a, \ph ] +i\stw \ps_a [\ps^a, \la ] \nn \\
 & - & \left. i2\stw \ps_a [\tc^a ,C ] + 
4[\ph ,B^{ai}][\la , B_{ai}] \right\}, \eea
and
\be
\cL_{-1}  =\frac{1}{\ep e^2}\tr \left\{ -4(D^a B_{ai})^2 - \f{1}{4} (F_{ab} 
+4i[B_{ai},B^i _{\ b}] )^2 \right\} . 
\ee 

Now, in sending $\ep$ to zero path integral localizes around the solutions
 of the following equations
\bea
 & & F_{ab} +4i[B_{ai},B^i _{\ b}] = 0  \nn \\ 
 & & D^a B_{ai}=0.  \label{BIA}
\eea

In appendix A we show that these equations imply 
\be
F_{ab}=B_{ai}=0,
\ee
and from $F_{ab}=0$ it follows that the instanton number vanishes. 
A flat 
connection on sphere can be  written globally as 
\[
A_a = g^{-1}\pl_a g
\]
for some gauge group element $g$. Therefore, the connection $A$ is
\[
A = A_i dx^i + (g^{-1}\pl_a g)dx^a\, .
\]
We gauge transform $A$ such that it lies in $\Sigma$ direction
\[
A \to gAg^{-1} + gdg^{-1}= g(A_i dx^i)g^{-1} + g(\pl_ig^{-1})dx^i =A'_i dx^i \, .
\]
Setting $A_a=0$ and $B_{ai}=0$,  $\cL_0$ greatly simplifies. However, because 
of the zero modes of  the operator $d_a$, one has to still keep the order $\ep$  
terms in $\cL_1$. We expand all fields in terms of eigen functions of $d_a$ 
and denote the zero modes by a $0$ superscript. Effectively we do the following 
 substitution
\[ \Phi(z,\zb ;w,\wb ) \to \Phi^0(z,\zb)+ \Phi(z,\zb ;w,\wb ) \]
where $\Phi(z,\zb ;w,\wb )$ on the RHS stands for the nonzero modes. The kinetic part 
of $\cL_0$ then reads 
\bea
\cL_{0\ {\rm kin}}  & = & \frac{1}{e^2}\tr \left \{ - \pl_a \la \pl^a \ph -( \pl_a C  
+\ep_{ab}\pl^b b)^2  -(\pl_a A_i)^2 \right. \nn \\
 & + & \left.4 \ps_{[i} \na_{a]} \ch^{ia} + 4\tc_{[i} \na_{a]} \tps^{ia} + 
\tc_a \na^a\ze - \ps_a \na^a\et + 4 \ps_{a} \na_{b} \ch^{ab}
 + 4\tc_{a} \na_{b} \tps^{ab} \right\}. \nn \\
\eea
Since $\ch_{ai}$ and $\tps_{ai}$ are self-dual and since there are no holomorphic 
one forms on sphere (see the Appendix), $ \cL_{0\ {\rm kin}}$ is nondegenerate.  
Thus in doing the integral over nonzero modes, one may drop the terms which are 
order of $\ep$. Keeping terms of order one, the integral over $\et, \ze ,\ch , 
\tps , \ps_i$ and $\tc_i$   
 results in a set of delta 
functions imposing 
the following constraints
\bea
& \na_a\ch^{ai}=0 , \ \na_a\tps^{ai}=0 \nn \\
& \na_a\ps^{a}=0, \ \ep^{ab}\na_a\ps_b =0  \nn \\
& \na_a\tc^{a}=0, \ \ep^{ab}\na_a\tc_b =0 . 
\eea
As was mentioned, these equations have no nontrivial 
solutions on sphere. Setting these fields to zero, $\cL_0$ reduces to
\[ \cL_{0}  = \frac{1}{e^2}\tr \left \{ - \pl_a \la \pl^a \ph -( \pl_a C)^2  
-(\pl_a b)^2  -(\pl_a A_i)^2 \right\} \]
where fields are all nonzero modes. 
Using the equation of motion for $A_i$ we obtain
\[ d^{\dagger}d A_i + {\rm terms\ proportional\ to\ \ep}=0 \]
as $A_i$ is a nonzero mode this equation implies that, up to $\ep$ order, 
$A_i=0$. The same happens for $\ph, b$ and $C$ fields. So in 
the limit $\ep \to 0$ all nonzero modes can be set to zero and one is 
left with a copy of $\cL_1$ in which fields now depend only on coordinates 
on $\Sigma$. From now on we call this reduced Lagrangian $\cL$ and drop the 
$0$ superscript on zero modes.

The reduced Lagrangian, $\cL$, which now describes a two-dimensional TFT, can be obtained by 
the BRST variation of $V$, where
\bea
V & = & \frac{1}{e^2}\int_{\Sigma} \tr \{ \ha \tc^i(\tH_i-2\stw D_iC
+\f{ 2\stw}{\gr} \ep_{ji}D^jb) +\ch(2H-2f -4i[b,C]) \nn \\
 & - &  \frac{1}{2\stw}\ \la(2D_i\psi^i +2\stw i[\tps ,b] + 
\stw i [\ze , C]) \},
\eea
and the BRST transformations of the two-dimensional fields are ($\de\equiv \{Q , \cdots\}$)
\bea \begin{array}{lllll}
 &\de A_i  =  -2\ps_i & \de b = \stw \tps   &\de C =  \frac{1}{\stw}\ze &  \nn \\ 
 &\de \ps_i =  -\stw i D_i \ph & \de \tps = -2[b ,\ph ] & \de \ze =-4[C,\ph ] &  \nn \\  
  &  \de \tc_i = i\tH_i & \de \ch = iH & \de \la = \stw \et & \de \ph =0 \nn \\
 &  \de \tH_i =2\stw i [\tc_i ,\ph ] & \de H = 2\stw i [\ch , \ph ]& \de \et 
= -2 [\la ,\ph ]. & \nn \\
 \end{array}
\eea
The fixed points around which path integral localizes are  those 
configurations that are BRST invariant. Thus, setting $\de \chi =H=0$ and 
$\de \tc_i=\tH_i=0$ and using the equation of motion for $H$ and $\tH_i$ we find the 
fixed point equations  
\bea
&& s=f+ 2i[b , C]=0 \nn \\
&& k= D_i C +\f{1}{\gr} \ep_{ij}D^jb=0 .\label{fix}
\eea
Squaring these equations implies that 
\bea
&& 0= \int \tr\, (\ha |s|^2 + |k|^2) \nn \\
&&\hspace{3mm} = \int \tr\, \{ \ha |f|^2 +2|[b, C]|^2 +2i f[b, C] 
+|D_iC|^2 + |D_ib|^2 + \f{2}{\gr}\ep_{ij}D^iCD^jb \} \, .\nn
\eea 
Using the definition of $f$ in (\ref{defi1}), we can see that the third and the 
last term cancel against each other. Therefore this integral is zero if and only if 
\bea 
&& f=0 \ ,\ [b , C]=0 \nn \\
&& D_iC=D_ib =0 .
\eea
Requiring that there are no reducible connections (as is the case for  
flat non-trivial $SO(3)$ bundles) it follows that the only solutions are $C=b=0$. 
 Therefore, following \cite{VW}, it can be seen that in this case the 
partition function is nothing but the Euler characteristic of the moduli 
space of flat connections over $\Sigma$.

\section{Perturbing by mass term}

The theory discussed so far does not have a mass gap \cite{WP}. To make the 
calculations more feasible we perturb the theory such that it has a mass gap.\footnote{ 
A similar perturbation has been considered in \cite{VW, LMASS} for $N=4$ SYM 
theory in four dimensions.}  
This enables us to integrate out most fields and reduce the path integral to 
a finite dimensional one.

\noindent 
The reduced 2-dimensional theory has a  $U(1)$ ghost number symmetry 
coming directly from the nonanomalous $U(1)$ symmetry of the  
underlying 4-dimensional $N=4$ SYM theory. Because of supersymmetry, the measure 
for nonzero modes is invariant under the $U(1)$ action. The ghost and the  
antighost zero modes, on the 
other hand, obey the same equations of motion such that there are equal number of 
ghost and antighost zero modes. This renders the measure to be invariant under the 
ghost symmetry of the action. Therefore the ghost symmetry is anomaly free. 

\noindent
As the measure is invariant under 
this symmetry, the correlation function of any operator that has a ghost 
charge is zero. Therefore,       
this symmetry allows us to perturb the 
Lagrangian, by adding  gauge invariant terms with  nonzero  ghost 
number, without changing the partition function. Thus, for example, since the mass 
term for the hypermultiplet (as we will see presently) 
consists of a term with negative ghost number and a 
term which is BRST exact, one expects that the partition function is invariant 
under  perturbing the Lagrangian by a mass term for the hypermultiplet.
 One can even go further to argue that an additional  mass term for the chiral 
multiplet $\Phi$ (which contains $\ph$ and $\la$) still leaves the partition function 
  invariant \cite{VW, LMASS}.  

In the following we are interested in the correlation functions of a set of 
BRST cohomology classes of the form 
\[ 
I(\vep)= \frac{1}{4\pi^2}\int_{\Sigma}\tr\left(\f{i}{\stw}\ph F +\f{1}{2} \ps\wedge\ps\right) 
+\frac{\vep}{32\pi^2}\int_{\Sigma}\tr\ph^2. 
\]
Part of this factor with an extra BRST exact term provides the mass for the chiral 
multiplet $\Phi$ \cite{WS}. The remaining part may have a nonvanishing expectation value in 
the mass deformed theory. This, in particular, implies 
that, in contrast to the partition function, the correlation functions of $I(\vep)$ 
(in the perturbed theory by mass for the hypermultiplet) may depend on the mass 
parameter.

The next problem is to  give a mass to $\ch ,\et ,\la ,\tps $ and 
$\ze$. This can be achieved  
by adding $V'$ and $V''$ to $V$, 
where 
\be V'= -\frac{2}{e^2}\int_\Sigma d\mu\ \tr \{ \ch \la \} \ee
\be V''=\frac{1}{e^2}\int_\Sigma d\mu\ \tr \{\tps C -\ha \ze b \}.
\ee 
To give a mass term to the bosonic fields $b, C$ and the fermionic one 
 $\tc_i$ we change 
the BRST transformation rules for $ \tH_i , \tps $ and $\ze$ to the 
following ones
\bea & & \de_m \tH_i =2\stw i [\tc_i ,\ph ] +\f{\stw}{\gr} m \ep_{ij}\tc^j \nn \\
& & \de_m \tps = -2[b,\ph] + imC \nn \\
& & \de_m \ze =-4[C,\ph ] - 2i mb . 
\eea 
Even though the metric is explicitly introduced via the  above first BRST transformation 
rule, note that the extra term is still invariant under metric rescaling ($\ep_{ij}
\sim g_1$).
%

Thus, in the following we will consider the theory defined by 
 the deformed action
\bea  S& = & I(\vep)+ i \de_m (V + t V' +\f{1}{2} \mb V'') \nn \\ 
&= &  I(\vep)+\frac{1}{e^2}
\int_{\Sigma}d\mu\ \tr \{  D_i \la 
D^i \ph  + D_i C D^i C 
+ D_i b D^i b - \f{2}{\gr}\ep^{ij} D_i bD_j C   \nn \\
 & + &\ha (f+2i[b,C] +t\la )^2 + 2i\stw t\ch \et - \ha |m|^2 C^2 
- \ha |m|^2 b^2\nn \\
 & + &\f{i \mb}{\stw}\ze\tps 
-\frac{m}{\sqrt {2 g_1}}\ep_{ij}\tc^i \tc^j 
+2i \mb \ph[b,C] -2i m\la[b,C] \nn \\
 & + &\f{ 4i}{\gr} \ep^{ij}\ps_i D_j \ch +\f{ 2i}{\gr}\ep^{ij} \tc_i D_j \tps + 
i\tc_i D^i\ze - i\ps_i D^i\et \nn \\
 & - & {\stw} \tps [\tps ,\la ] +4\stw \ch
[\ch ,\ph] -  4\stw \tps [\ch ,C]+ 2\stw \ch
[\ze ,b]\nn \\
 & + & {\stw} \tps [\et ,b] - \f{2\stw}{\gr} \ep^{ij} \ps_i 
[\tc_j ,b]  + \stw \tc_i [\tc^i , \ph ] - \stw \ps_i [\ps^i , \la ]  \nn \\ 
& + & 2\stw \ps_i [\tc^i ,C ]  - \frac{1}{2\stw }\ze [\ze , \la] 
+  \frac{1}{\stw } \ze [\et , C]
  -  2[\ph ,b][\la , b] - 2 [\ph ,C][\la ,C]
  \} .  \label{deformed}
\eea
Notice that although the new BRST charge does not square to a gauge 
transformation (because of those new terms proportional to $m$), Lagrangian 
 remains  BRST invariant. This can be understood if we notice that $\de_m^2$ acting 
on fields  generates (up to a gauge transformation) a $U(1)$ action. Let 
$\de_T \equiv \f{1}{i\stw m}\de_m^2$ and $\bet \equiv b +iC ,\  
\psi \equiv \tps +\f{i}{2} \ze$, then $U(1)$ group acts as 
\bea
&& \de_T \bet = -i \bet \ \ ,\ \ \de_T\psi =-i \psi \nn \\
&& \de_T \tc_i =\f{1}{\gr}\ep_{ij}\tc^j \ \ ,\ \ \de_T \tH_i =\f{1}{\gr}\ep_{ij}\tH^j 
\nn
\eea
thus the fields  $\bet ,\ \psi ,\ \tc_\zb$ and $\tH_\zb$ all have charge $-1$, with 
their complex conjugate having charge $+1$. All other fields have zero  charge 
under this $U(1)$ group. The fact that $S$ is invariant under 
$\de_m$ then follows since $V , V'$ and $V''$ all have zero $U(1)$ charge.

Before continuing the analysis, it is important to understand the relation 
between the perturbed and unperturbed theories. 
Since the perturbing terms proportional to $t$ and 
$\mb$ are BRST exact, one may expect  that correlation functions are going 
to be independent of these 
two parameters, but actually this is not  true in general: 
adding  $\de_m V'$ and $\de_m V''$ to 
the Lagrangian may result in 
some new set of fixed points flowing in from infinity and  deforming the original moduli 
space of solutions \cite{WRE} such that the path integral gets contribution from these new 
fixed points. The theory will be  independent of $t$ and $\mb$ if in varying 
these parameters Lagrangian remains  nondegenerate and  the perturbation 
does not introduce new components to the moduli space of fixed points. 

We first discuss the situation for $t=0$ with arbitrary $m$ and $\mb$. The perturbed 
Lagrangian (with $t=0$) can also be derived upon  reducing   
the $N=4$ theory 
broken to $N=2$ by the mass term. Had we started with $N=2$ theory 
with one massive hypermultiplet 
in the adjoint representation of the gauge group in four dimensions, 
we would have ended up with the same 
above perturbed Lagrangian  after reduction.

The fixed 
point equations are those of (\ref{fix})  
together with (setting 
$\de_m \tps=\de_m \ze =\de_m \et =\de_m \ps_i=0$) 
\bea
 [\bet , \ph]= \ha m\bet  \ ,\ \ 
 [\la , \ph]=0 \ ,\ D_i \ph=0 . \label{beta}
\eea 
If $\ph$ is not identically zero then, being covariantly constant, 
it never vanishes and, in particular,  
can be diagonalized globally such that the bundle $E$   
splits  as a sum of  line bundles \cite{FU}. 
Moreover, if $\bet \neq 0$, the first equation in (\ref{beta}) fixes 
$\ph$ (up to a sign)     
\be
\ph = \f{m}{4} \lf(
\ba{lc}
1 & 0 \\
0 & -1
\ea
\ri) \label{global}
\ee
with $\bet$ as 
\be
\bet = {\tilde \beta}\lf(
\ba{lc}
0 & 0 \\
1 & 0
\ea
\ri). 
\ee
Now the equations (\ref{fix}) become
\bea
&& {\tilde f} + 2|{\tilde \beta}|^2 =0 \nn \\
&& {\bar D}{\tilde \beta}=(\del_{\zb} -i A_{\zb}){\tilde \beta} =0 \nn
\eea
(notice $f=\ha {\tilde f}\si_3$, where ${\tilde f}$ here is  the $U(1)$ curvature). 
Note that $\ph =\qu m \si_3$ 
corresponds to a point, in the classical moduli space of vacua, where a component of 
the hypermultiplet becomes massless.\footnote{ As eq. (\ref{beta}) fixes $\ph$ up to 
a sign, there are indeed two such singular points in the classical moduli space of vacua.} 
 The relevant fixed points  
   are then 
determined by the above equations. Clearly one 
can then argue that the path integral over massless modes computes the 
Euler characteristic of the moduli space of $U(1)$ flat connections. However, 
to evaluate the contribution of this singular point to the path integral,  
one still has to do the  integral over the massive modes.
 
\noindent
This is not an easy task, but there is a special case where this  point 
($\ph =\qu m\si_3$)  
does not make any contribution. This occurs upon restricting to the  nontrivial $SO(3)$ 
bundles. As discussed above, a nonzero  $\ph $ breaks the gauge group down to $U(1)$. 
In particular,  $SO(3)$ bundles split as 
\be
E = L  \oplus {\cal O}\oplus  L^{-1}, \label{split} 
\ee
where $L$ is the $U(1)$ line bundle and ${\cal O}$ is a trivial line bundle. 
In this case, $w_2(E)$, which measures the nontriviality of the bundle $E$, 
turns out to be the mod two reduction of $c_1(L)$, the first Chern class of $L$
 \cite{VW}. Thus if $f=0$, as is required by eqs. (\ref{fix}), $w_2(E)$ has 
to be zero --    
implying that flat nontrivial $SO(3)$ bundles do not admit  reducible connections.
 Therefore, in this case, the point $\ph=\qu m \si_3$ does 
not contribute to the path integral.

Let now discuss the case that $t\neq 0$. 
The fixed point equations (\ref{fix}) turn into the following 
equations  
 ($\bet \equiv b+i C$  
with $\ep_{z\zb}=i{\sqrt g_1}g_{z\zb}$)
\bea
&& f + [\beb , \bet] + t\la =0 \nn \\
&& {\bar D}\bet =0 \ \ ,\ \ {D}\beb =0 .\label{perturb}
\eea
The  vanishing argument now fails; $f=\bet=0$ (and $\la =0 $)  are not the 
only solutions,   there are  new fixed points with $f\neq 0$ 
 contributing to the partition function. Since the connection 
is not bounded to be flat 
any more,   
a  set of $U(1)$ connections, in all classes of $U(1)$ bundles, 
appear in the moduli space of 
solutions. Moreover,   
the point $\ph=\qu m \si_3$ may contribute to 
the path integral even for 
nontrivial bundles.   In the following we 
single out  this point  from our 
discussion and treat it independently.

\vspace{4 mm} 
\noindent  
\underline {Integrating $\la$, $\et$ and $\chi$}  

Perturbing by $V'$ now allows us to integrate out the fields 
$\la$, $\et$ and $\chi$. 
Using the equations of motion for $\la$ and $\et$ we get
\bea 
t^2\la & = &  D^2\ph -t(f+2i[b,C]) +2i m[b,C] + \stw [
\ps_i ,\ps^i] \nn \\
& + & \stw [\tps ,\tps ] +\frac{1}{2\stw }[\ze ,\ze]+ 2[b,[\ph,b]]
+2 [C,[\ph ,C]] 
\eea
and
\[ 
\ch = \frac{1}{2\stw t}\left\{ -D_i\ps^i +i\stw[b, \tps] +
\frac{i}{\stw}[C,\ze ] \right\}. 
\]
Putting these back into the Lagrangian yields
\bea  S\!\!\! & = &\!\!\!I(\vep)+ \! \frac{1}{e^2}\int_{\Sigma}d\mu\ \tr 
\!\!\left\{  D_i C D^i C + D_i b 
D^i b - \f{2}{\gr}\ep^{ij} D_i bD_j C + \f{2i}{\gr}\ep^{ij} \tc_i D_j \tps +
i\tc_i D^i\ze \right. \nn \\
& - &
\!\!\! \f{2\stw}{\gr} \ep^{ij} \ps_i [\tc_j ,b]  + 2\stw \ps_i [\tc^i ,C ] 
 + \stw \tc_i [\tc^i , \ph ] 
  - \ha |m|^2C^2 -\ha |m|^2 b^2 +\f{i\mb}{\stw}\ze\tps  \nn \\
& - &
\frac{m}{\sqrt {2 g_1}}\ep_{ij}\tc^i \tc^j + 2i \mb \ph[b,C] + \frac{1}{t} 
{\mbox { \Large \{ }} (f+2i[b,C])  \nn \\
&\times  &
\! \left( D^2\ph +2i m[b,C] + \stw [
\ps_i ,\ps^i]+ \stw [\tps ,\tps ] +\frac{1}{2\stw }[\ze ,\ze]
+ 2[b,[\ph,b]]+2 [C,[\ph ,C]]\right)  \nn \\
& + &
\! \left. \frac{i}{2\stw}\left(-D_i\ps^i 
+i\stw[b, \tps] +\frac{i}{\stw}[C,\ze ]\right)\left(\f{-4}{\gr}\ep^{kl}D_k\ps_l
 -i4\stw[C, \tps] -2i\stw[\ze ,b] \right) \right\} \nn \\
& + &
\!\!\! \frac{1}{2t^2} \left\{\!\!\left(D^2\ph +2i m[b,C] + \stw ([
\ps_i ,\ps^i]+ [\tps ,\tps ] +\frac{1}{4 }[\ze ,\ze])
+ 2[b,[\ph,b]]+2 [C,[\ph ,C]]\right)^2 \right. \nn \\
& + &
\! \left.\left. \!\!\! \stw
\left(-D_i\ps^i +i\stw ([b, \tps] +\frac{1}{2}[C,\ze ])\right)
\left[\left(-D_l\ps^l 
+i\stw ([b, \tps] +\frac{1}{2}[C,\ze ])\right),\ph \right] 
\right\}\right\}. \label{effective}
\eea
Terms proportional to $1/t$ are indeed BRST trivial, and can be written
\[ 
\frac{i}{\stw t} \de_m \left\{\left(f+2i[b,C]\right)\left
(-D_i\ps^i +i\stw[b, \tps] +
\frac{i}{\stw}[C,\ze ]\right) \right\}.
\]
Terms proportional to $1/t^2$ are also combining into 
\bea & & \frac{i}{2\stw t^2} \de_m \left\{\left(-D_i\ps^i +i\stw[b, \tps] +
\frac{i}{\stw}[C,\ze ]\right) \times \right. \nn \\
 & & 
\left. \left(D^2\ph +2i m[b,C]+ \stw [
\ps_l ,\ps^l]+ \stw [\tps ,\tps ] +\frac{1}{2\stw }[\ze ,\ze]
+ 2[b,[\ph,b]]+2 [C,[\ph ,C]]\right) \right\}. \nn
\eea
In the effective Lagrangian (\ref{effective}), the kinetic terms are 
nondegenerate for all values of $t$ and since those terms proportional 
to $t$ are still in a BRST exact form, the path integral does not 
depend on $t$.    
\vspace{4 mm}

\noindent
\underline {Large $t$ Limit and The Integration over $b$, $C$, $\ze$, 
$\tps$ }

As argued above, for nontrivial $SO(3)$ 
bundles the point $\ph =\qu m\si_3$ does not contribute.

For $t\neq 0$, because of 
the supersymmetry, even after integrating out $\la , \et$ and $\ch$  the 
singularity still persists at  $\tr \ph^2= \f{1}{8} m^2$. 
As we have chosen $\ph$ to be a real scalar field, reality of the action requires 
that $m$  to be a real parameter.
 However, to regulate the contribution of the points in the neighborhood of 
$\tr \ph^2= \f{1}{8}m^2$, we allow $m$ to 
have a small imaginary part.   
If there is going to be  any singularity when $\ph$ approaches 
$m$, it has to show up in the final result when we take the limit Im $ m\to 0$. 
This can be thought of as a kind of regularization by analytic continuation.

Now let us consider the large limit of $t $. Since the 
kinetic terms remain  nondegenerate  we can actually take 
$t \to \infty $.
Using the auxiliary field $\tH_i$, in this limit we are left with the action
\bea
S & = & \frac{1}{e^2}\int_{\Sigma}d\mu\ \tr \left\{ 
- \ha \tH^i(\tH_i-2\stw D_iC
+\f{ 2\stw}{\gr} \ep_{ji}D^jb)+ \f{2i}{\gr}\ep^{ij} \tc_i D_j \tps +
i\tc_i D^i\ze \right. \nn \\
 & - &  \ha |m|^2 C^2 - \ha |m|^2 b^2 +\f{i\mb}{\stw}\ze\tps 
-\frac{m}{\sqrt {2 g_1}}\ep_{ij}\tc^i \tc^j 
+2i \mb \ph [b,C]  \nn \\
 & - & \left. \f{ 2\stw}{\gr} \ep^{ij} \ps_i [\tc_j ,b] + 2\stw \ps_i [\tc^i ,C ] 
 + \stw \tc_i [\tc^i , \ph ] \right\}+ I(\vep). \nn 
\eea
 $\cL $ can still be written as a sum of 
BRST exact term
\[  
i\de_m \left\{\frac{1}{e^2} \tr \{ \ha \tc^i(\tH_i-2\stw D_iC
+\f{ 2\stw}{\gr} \ep_{ji}D^jb)+ \f{1}{2} \mb( \tps C -\ha \ze b) \}\right\}
\]
and $I(\vep)$.
The integral over $C$ 
gives a factor of \mbox{ $\left(\det (\f{1}{2e^2} |m|^2)\right)^{-\f{1}{2}}$}
 and leaves 
\bea
S & = & \frac{1}{e^2}\int_{\Sigma}\tr \left\{-\ha\tH^i\tH_i +\stw  \tc^i[\tc_i ,\ph]
-\frac{m}{\sqrt {2 g_1}}\ep_{ij}\tc^i\tc^j+ \f{2i}{\gr}\ep^{ij} \tc_i D_j \tps +
i\tc_i D^i\ze \right. \nn \\
 & + &\frac{1}{|m|^2}(D_i\tH^i - 2[\tc_i ,\ps^i])^2 -2\frac{\mb}{m}[b,\ph]^2 
+\frac{2i\stw }{m}[b,\ph](D_i \tH^i  - 2[\tc_i ,\ps^i])   
 \nn \\
 & - & \left. \f{\stw}{\gr} \ep_{ji}\tH^i D^jb - \ha |m|^2 b^2 
+\f{i\mb}{\stw}\ze\tps   - \f{2\stw}{\gr} \ep^{ij} \ps_i [\tc_j ,b]  
 \right\}+ I(\vep). \nn 
\eea
Next we would like to integrate out $b, \ze$ and  $\tps$. It is easy 
to integrate out $\ze$ and $ \tps$ using their equations of motion.
 In the evaluation of 
determinants, which appear in doing the integral over $b$ and finally 
over $\ch_i$, we always assume that $\ph$ is a  constant field. This can
be justified finally when the integral over the gauge fields constrains  
$\ph$ to be 
constant. The equation of motion for 
$b$  yields
\be
b^A  =\frac{\stw}{|m|^2} K^{AB}\left(-\f{1}{\gr}\ep^{ij}D_j\tH_{i} +\f{2}{\gr}
\ep^{ij}[\tc_i ,\ps_j]
 - \frac{2i}{ m}
[(D_i\tH^i - 2[\tc_i ,\ps^i]),\ph]\right)^B , 
\ee
where we have defined ($A$ and $ B$ are lie algebra indices)
\[ 
{ K}^{AB}\equiv (1 - \frac{8}{m^2}\tr \ph^2)^{-1}(\de^{AB}-\frac{8}{m^2}\ph^A\ph^B).
\]
Replacing $b$ in the action,  
 we obtain
\bea
S\!\! & = &  I(\vep) +  \frac{1}{e^2}\int_{\Sigma} d\mu\ \tr \left\{-\f{1}{2}\tH^i\tH_i 
+\stw  
\tc^i[\tc_i ,\ph] \ri. \nn \\ 
& - & 
\lf. \frac{m}{\sqrt {2g_1}}\ep_{ij}(\tc^i\tc^j - \frac{4i}{|m|^2}D_l \tc^l 
D^i\tc^j)
 +  \frac{1}{|m|^2}\!(D_i\tH^i - 2[\tc_i ,\ps^i])^2 \right\} \nn \\
& + &\!\!\!\!\frac{1}{|m|^2}\left(-\f{1}{\gr}\ep^{ij}D_j\tH_{i} +\f{2}{\gr}
\ep^{ij}[\tc_i ,\ps_j] 
- \frac{2i}{m}
[(D_i\tH^i - 2[\tc_i ,\ps^i]),\ph] \right)^A K^{AB} \nn \\
& \times & \left(-\f{1}{\gr}\ep^{kl}D_l\tH_{k} +\f{2}{\gr}\ep^{kl}[\tc_k ,\ps_l] 
-\frac{2i}{m}
[(D_l\tH^l  - 2[\tc_l ,\ps^l]),\ph] \right)^B \  , \nn
\eea
and a factor of 
\[
\lf(\det (\frac{1}{\stw e^2}\mb)
(\det \f{1}{2e^2}|m|^2)^{-1}
(\det  (1 + \frac{8}{m^2}({\rm ad} \ph)^2))^{-\ha}\ri)
_{\Omega^0 \otimes E}\, ,
\]
where $({\rm ad}\ph)_{AB}= -f_{ABC}\ph_C$ and 
$\Omega^0$ indicates  the space of zero-forms.
 
\noindent
The following are easily derived, 
\bea 
\de_m \left\{(D_l \tc^l)(D_i\tH^i -2[\tc_i ,\ps^i])\right\} &=&  
i(D_i\tH^i -2[\tc_i ,\ps^i])^2 -2\stw i (D_i \tc^i)[D_l \tc^l,\ph] \nn \\
&+& 
\f{\stw m}{\gr}
\ep^{ij} D_i\tc_j D_l \tc^l \nn \ ,
\eea
and 
\bea
\de_m \left\{\f{1}{\gr}\ep^{ij} D_j\tc_i +\frac{2i}{m}[D_i \tc^i,\ph] \right\}\!\!\! 
&=& 
\!\!\!\f{i\ep^{ij}}{\gr} D_j\tH_i -\f{2i}{\gr}\ep^{ij}[\tc_i ,\ps_j]- \frac{2}{m}
[(D_i\tH^i -2
[\tc_i ,\ps^i]),\ph] \nn \\
\de_m^2 \left\{\f{1}{\gr}\ep^{ij} D_j\tc_i +\frac{2i}{m}[D_i \tc^i,\ph] \right\}\!\!\!
 &=&\!\!\!
i\stw m D_l \tc^l -\frac{4i\stw}{m}\left[[D_l \tc^l,\ph],\ph \right]. \nn
\eea
Using these, the action can be written as
\bea 
S   &=& I(\vep)+ \frac{1}{e^2}\int_{\Sigma}d\mu\ \tr \left(-\ha\tH^i\tH_i +
\stw  \tc^i[\tc_i ,\ph]
-\frac{m}{\sqrt {2g_1}}\ep_{ij}\tc^i\tc^j \right) \nn \\
&-& 
 \frac{i}{e^2 |m|^2}\int_{\Sigma}d\mu\ \  \de_m \left\{\f{}{}\tr 
\left((D_l \tc^l)(D_i\tH^i -2
[\tc_i ,\ps^i])\right)  
 \right.  \\
& + &  \left(\f{\ep^{ij}}{\gr} D_j\tc_i +\frac{2i}{m}[D_i \tc^i,\ph]
\right)^A 
 K^{AB} \nn \\
& \times & \left.
\left(\f{\ep^{kl}}{\gr}( D_l\tH_k -{2} 
[\tc_k ,\ps_l]) +\frac{2i}{m}[(D_l\tH^l -2[\tc_l ,\ps^l]),\ph]\right)^B 
 \nn \right\}.
\eea
Note that the integration over  $b$, $C$, $\ze$ and  
$\tps$ has not destroyed the manifest BRST exactness of the action, in particular,  
the variation of $S$ with respect to  $\mb$ is still a BRST commutator. 
\vspace{4 mm}

\noindent
\underline {Large $\mb$ Limit and The Final Reduction}

We note the partition function is formally independent of $\mb$ (since  
 the variation of the partition function with respect to $\mb$ gives an BRST 
exact expression) and is really 
independent of $\mb$ if in varying $\mb$ the Lagrangian remains  
nondegenerate with a good behaviour at infinity in field space. The mass 
term for $\tc_i$, the term $\tH^i\tH_i$, and  the form of the cohomology 
classes that we have added by hand, guarantee that this is actually the case. 
Having this freedom in the value of $\mb$, we simply set $\mb=\infty $. 
This leaves us with the action 
\[S =I(\vep)+ \frac{1}{e^2}\int_{\Sigma}d\mu \left\{-\ha\tH^{iA}\tH_i^{\ A}- \tc^{iA}
\left(\frac{1}{\sqrt {2g_1}}m
\ep_{ij}\de_{AB} - 2 if_{ABC}\ph_C g_{ij}
\right)\tc^{jB} \right\},
\]
and the partition function reads 
\[ Z[\vep,m] = \int \cD(A_i,\ps_i, \ph ,\tH_i ,\tc_i)\left(\frac{\det (\frac{1}{\stw e^2}
\mb)}{(\det \f{1}{2e^2}|m|^2)(\det  (1 + \frac{8}{m^2}({\rm ad} \ph)^2))^{\ha}}\right)
_{\Omega^0\otimes E}
 e^{-S}, \]  
The  explicit appearance of $m$ on the LHS reminds us that, although independent of 
$\mb$, $Z$  does depend on $m$. This is so because  
$m$ was  introduced  through the BRST transformation laws. This is reminiscent of 
holomorphicity of $N=1$ theories in four dimensions.

%
%

Doing the integral over $\tc^i$ gives a similar determinant, but this time  
over the space of one-forms.
 Putting all pieces
 together one gets  
\be
Z[\vep,m] = \int \cD(A_i,\ps_i, \ph)\left( \frac{\left [\det m(1 
- \frac{2i\stw}{m}{\rm ad} \ph)\right ]
_{\Omega^1\otimes E}}{\left [\det m
(1 - \frac{2i\stw}{m}{\rm ad} \ph)\right ]_{\Omega^0\otimes E}}\right)
e^{ \lf(\frac{-1}{4\pi^2}\int_{\Sigma}\tr\left(\f{i}{\stw}\ph F +\f{1}{2} \ps\wedge\ps\right) 
-\frac{\vep}{32\pi^2}\int_{\Sigma}\tr\ph^2 \ri)}  . \label{pf}
\ee
Notice that, as expected,  $\mb$ cancels out between the fermionic and bosonic 
determinants. The  integral over $\ps_i$ provides a symplectic measure for the 
gauge fields $A_i$ \cite{WRE}. Performing the path integral over $\ph$ and $A_i$ is 
now straightforward.  In appendix B,  
using the Faddeev-Popov gauge fixing technique,   
it has been shown that the integral over the gauge fields constrains $\ph$ to be constant 
and hence the path integral calculation reduces   
 to a finite 
dimensional integral over constant $\ph$ \cite{BTLE}. Explicitly, for $SO(3)$ gauge group we have 
\bea
Z[\vep,m] &=&  m^{3(g-1)} \sum_{n\in {\bf Z}} \int d\ph\ \ph^{2-2g}
(1 - \frac{8}{m^2} \ph^2)^{g-1}\lf(\f{m-2\stw \ph}{m+2\stw\ph}\ri)
^{2n + 1} \nn \\ 
&\times & \exp \left( -i\stw\frac{\ph (2n + 1)}{4\pi}-\frac{\vep \ph^2}{32\pi^2} \right). \label{Z[m]}
\eea


\section{Discussion} 

We have reduced the calculation of the correlation functions in the mass 
deformed theory to a finite dimensional integral in (\ref{Z[m]}). We can now  
perform the  
sum over $n$ which results in a delta function restricting $\ph$ to obey the 
following equation
\be
\exp \lf( \f{i\stw\ph}{2\pi}\ri) =  \lf(\f{m-2\stw \ph}{m+2\stw\ph}\ri)^2\, .\label{BET}
\ee
Therefore 
\be
Z[\vep,m] =m^{3(g-1)}\sum_{\ph_s} \ph_s^{2-2g}(1 - \frac{8}{m^2} 
\ph_s^2)^{g-1}\lf(\f{m-2\stw \ph_s}{m+2\stw\ph_s}\ri)\exp \left( -i\frac{\stw\ph_s}{4\pi}
-\frac{\vep \ph_s^2}{32\pi^2}\ri)\, ,
\ee
where $\ph_s$ is a solution to the eq. (\ref{BET}). A similar result for the correlation functions 
of a topological field theory corresponding to the Hitchin equations has 
been derived in \cite{NEK2}.

To this one still has to add the contribution of the point $\ph=\qu m\si_3$. However, 
note that from the discussion we had in section 3, for nontrivial $SO(3)$ bundles, 
this point contributes only if we perturb to 
$t\neq 0$.\footnote{ As is discussed in \cite{WRE} the contribution of the original moduli 
space is invariant under perturbing to $t\neq 0$.}  
Thus if we are interested in the limit of $t=0$, we can just ignore the 
contribution of this point. 


In conclusion we note two observations. Firstly, the result is $m$-dependent as 
might be expected from the discussion in section 3. Note  in particular that the 
 expression (\ref{Z[m]}) has the right 
behavior when $m \to \infty$; in this limit, the solutions of eq. (\ref{BET}) 
are reduced to 
\[
\ph_s = 2\stw \pi^2 l
\]
and therefore eq. (\ref{Z[m]}) reduces to the expression for the corresponding correlation 
functions in the say pure $N=2$ theory \cite{WRE}.  
The extra factor, $m^{3(g-1)}$, is left from the integration 
over the heavy fields in that limit. The power of $m$ is in accord with the dimension of 
the moduli space of flat connections which is 
\[
{\rm dim}{\cal M}= 6g -6\, .
\]
Any two zero modes of $\ch_i$ are absorbed by the corresponding mass term in the Lagrangian 
and gives a power of $m$.

Secondly, we recall that, in general, $S$-duality relates the strong and weak couplings and 
swaps the gauge group with its dual group. However, as in the limit where 
$S^2$ shrinks only instantons with $k=0$ contribute to the path integral, unlike \cite{VW},   
the correlators in the effective theory do not depend 
on the modular parameter ``$\tau$''. 
Hence the action of S-duality is now simply
to exchange the gauge group $SU(2)$ with $SO(3)$. Thus to derive that $S$-duality holds in 
this calculations, we must extend it for the $SU(2)$ case; in particular, 
the contribution of the point $\ph =\qu m \si_3$ must be taken into account. 
Amusingly, one can infer properties of this contribution by demanding $S$-duality.


\section*{Acknowledgements}

I am very grateful to Jim McCarthy for all his support 
and assistance throughout this work. I would also like to thank  
Nicholas Buchdahl for useful discussions.

\appendix

\pagebreak

\section{The Vanishing Argument}
In this appendix we want to discuss the solutions to eqs. (\ref{BIA}):
\bea 
 & &k= F_{ab}+4i [B_{ai} , B^{i}_{\ b}]  = 0  \nn \\ 
 & &s= D^a B_{ai}=0. 
\label{BIAS}\nn 
\eea
Let first analyze the second equation. After squaring we get 
\bea 
\int \tr (D^a B_{ai})^2 & = &  - \int \tr B^{ai}( D_aD_b B^b_{\ i}) \nn \\
 & = & - \int \tr \left(B^{ai}D_bD_a B^b_{\ i} + B^{ai}[D_a ,D_b] 
B^b_{\ i} \right) \nn \\
 & = &  \int \tr \lf((D_a B^b_{\ i})(D_b B^{a i}) + R_{ab}B^{ai}B^b_{\ i} 
-iB^{ai} [F_{ab} ,B^b_{\ i}]  \ri) \nn \\
 & = &  \int \tr \lf((D_a B^b_{\ i} +D^b B_{a i}- D^b B_{a i} )(D_b B^{a i})
 + \ha R\ B^{ai}B_{a i} -iB^{ai} [F_{ab} ,B^b_{\ i}] \ri) \nn \\
 & = & \int \tr\lf( (D_b B_{a i})^2 -\ha (D_{[a} B_{b]i})^2 + \ha R\ B^{ai}B_{a i}
-iB^{ai} [F_{ab} ,B^b_{\ i}] \ri) \label{square}
\eea
where we used the fact that in two dimensions, Ricci tensor takes a simple form
\[ R_{ab}= \ha g_{ab}R \]
and
\bea & & [D_a,D_b]B^{ci}= R^c_{\ dab}B^{di} +i [F_{ab}, B^{ci}] \nn \\
 & & [D_a,D_b]B^{ai}= R_{ab}B^{ai}+i [F_{ab}, B^{ai}].
\eea  
Since $B_{\mu\nu}$ is  self-dual, we have 
$B_{w\zb}=B_{\wb z}=0$,  
hence 
\bea 
(D_{[a} B_{b]i})(D^{[a} B^{b]i}) & = & (D_{\wb} B_{wz})(D^{\wb} B^{wz})+
 (D_w B_{\wb\zb})(D^w B^{\wb\zb}) 
\nn \\ 
& = & (D^w B_{wz})(D_w B^{wz})+
 (D^{\wb} B_{\wb\zb})(D_{\wb} B^{\wb\zb}) \nn \\ 
& = & (D^a B_{ai})(D_b B^{bi}).  
\eea 
Putting this back into (\ref{square}) we get
\be
\tht \int \tr (D^a B_{ai})^2  = 
\int \tr\lf( (D_b B_{a i})^2 + \ha R\ B^{ai}B_{a i}
-iB^{ai} [F_{ab} ,B^b_{\ i}] \ri). 
\ee 
Upon adding the squares of the sections $k$ and $s$, we have 
\bea
\int \tr(\qu k^2 +3 s^2) &=& \int \tr\lf\{ \qu (F_{ab})^2 -4[B_{ai} , B^{i}_{\ b}]^2 
+2iF_{ab}[B^{ai} , B_{i}^{\ b}] +2(D_b B_{a i})^2 \ri.
\nn \\
&+& \lf. R\ B^{ai}B_{a i}
-2iB^{ai} [F_{ab} ,B^b_{\ i}] \ri\}
\nn \\
&=&
 \int \tr\lf\{ \qu (F_{ab})^2 -4[B_{ai} , B^{i}_{\ b}]^2 
 +2(D_b B_{a i})^2 + R\ B^{ai}B_{a i}
\ri\} \nn
\eea
 the right hand side vanishes if and only if $k=s=0$. However, 
for sphere ($R>0$) all terms on the RHS are positive definite so a solution  
to $k=s=0$ has necessarily  $B^{ai}=0$.   
This leaves us with  the equation
\[
F_{ab}=0
\]
this equation implies that the connection is locally a pure 
gauge $A_a=u^{-1}d_a u $ for some $SU(2)$ matrix $u$. However, 
as the transition functions for  $SU(2)$ bundles on  sphere 
are trivial, the connection can be written globally as a pure gauge and be 
gauged away. Moreover, one can argue that this can be done continuously
 all over $\Sigma$. Thus we can set $A_a=0$ everywhere. 

\noindent
More 
rigorously if $\{U_\al\}$ is an open covering of $\Sigma$ by contractible sets 
and $\{V_i\}$ is an open covering of $S^2$ by such  sets, the sets $U_\al\times 
V_i$ give an open cover of $\Sigma\times S^2$ by contractible sets. On the 
intersection of two patches, the connection 
$A$ now satisfies 
\[
A_{\al i}= g^{-1}_{\al i \bet j}A_{\bet j}g_{\al i \bet j} +g^{-1}_{\al i \bet j}d
g_{\al i \bet j},
\]
or 
\[
d g_{\al i \bet j}+ A_{\bet j}g_{\al i \bet j} - g_{\al i \bet j}A_{\al i}=0 . 
\]
Since the $S^2$ component of the curvature is zero we have that 
$(A_a)_{\al i}= u^{-1}_{\al i} d_a u_{\al i}$. Putting this in the above equation 
yields
\[
d_a (u_{\al i}g_{\al i \bet j}u^{-1}_{\bet  j})=0 .
\]
Therefore $\gb_{\al i \bet j}\equiv u_{\al i}g_{\al i \bet j}u^{-1}_{\bet  j}$ does 
not depend on the coordinates of $S^2$. This implies that $\gb_{\al i \bet j}$'s are 
a set of locally constant  
transition functions equivalent to $g_{\al i \bet j}$ and for a fixed point on 
$\Sigma$ define a map from $S^1$ to $SU(2)$. This map  is trivial 
so $\gb_{\al i \al j}$ belongs to the  conjugacy class of identity
\[
\gb_{\al i \al j}=\gb_{\al i}\gb^{-1}_{\al j}=u_{\al i}g_{\al i \al j}u^{-1}_{\al  j}
\]
or $(\gb^{-1}_{\al i}u_{\al i})g_{\al i \al j}(\gb^{-1}_{\al j}u_{\al j})^{-1}=1$. 
Now consider $(\gb^{-1}_{\al i}u_{\al i})g_{\al i \bet j}(\gb^{-1}_{\bet j}u_{\bet j})^{-1}$. 
This is a constant matrix in the $S^2$ direction. Since $g_{\al i \bet j}=
g_{\al i \bet i}g_{\bet i \bet j}$ it is equal to $(\gb^{-1}_{\al i}u_{\al i})
g_{\al i \bet i}(\gb^{-1}_{\bet i}u_{\bet i})^{-1}$, and since  $g_{\al i \bet j}=
g_{\al i \al j}g_{\al j \bet j}$ it is equal to $(\gb^{-1}_{\al j}u_{\al j})
g_{\al j \bet j}(\gb^{-1}_{\bet j}u_{\bet j})^{-1}$. Thus it is in fact independent 
of the index $i$ and  therefore defines a matrix $\tilde{g}_{\al \bet}$ depending 
only on $x\in U_{\al \bet}$ and satisfying the cocycle condition.\footnote{ The 
proof of this part was provided  by 
Nicholas Buchdahl. }  

Since the  transition functions are independent of $i$, 
therefore $(A_{\Sigma})_{\al i}$ do not depend on $i$ index and $A_a$ can 
be gauged away.

It is now easy to see that the  flatness condition, $F_{ab}=0$, necessarily 
requires  the 
instanton number to be zero. The curvature locally takes the form 
\[
F=dA + A \wedge A
\]
therefore {\em locally} we can write
\[
\tr( F\wedge F) = d \ \tr (A\wedge dA + {\mbox{$\frac{2}{3}$}} A\wedge A \wedge A),
\]
but since $A_a=0$, instanton number reads 
\[
k= \f{1}{8\pi^2}\int_{\Sigma \times S^2}\tr F\wedge F = \f{1}{8\pi^2}
\int_{\Sigma \times S^2} d_{C} 
\ \tr (A_{\Sigma}\wedge d_C A_{\Sigma})
\]
where the subindex $C$ indicates differentiating with respect to the coordinates on $S^2$. 
Note that the integrand is still a local one. However, we showed that the transition functions 
are independent of the local coordinates on $S^2$. Therefore, for a fixed point on 
$\Sigma$, $A_{\Sigma}$ is globally defined on $S^2$. This means that the  
integral over $S^2$ is a total divergence and gives zero for the instanton number. 
In summary, we have learned that if the bundle $E$ admits a flat connection in $S^2$ direction then 
it has to be trivial (for those  bundles that are classified only by instanton 
number) and $k$, the instanton number, is zero.

\pagebreak

\section{Faddeev-Popov gauge fixing} 

In this appendix we want to show how the  eq. (\ref{Z[m]}) is obtained  
starting from (\ref{pf}). 
To evaluate the path integral over gauge fields and $\ph$, following 
\cite{BTLE}, we 
choose the so called unitary gauge in which one rotates the lie algebra valued field 
$\ph^a$ to the Cartan subalgebra by conjugation, i.e.\ we choose $\ph_{\pm}=0$, where 
\[ \ph = \ph_3 \ta_3 + \ph_+ \ta_+ + \ph_- \ta_-. \]
This gauge can always be achieved at least locally, but there might be some topological 
obstruction to impose it globally \cite{BTLE}. Implementing this gauge in the path integral 
requires to introduce the Faddeev-Popov ghosts $c$ and antighosts ${\bar c}$ together with 
a bosonic auxiliary field $b$. These fields transform under a BRST operator $\de$ like 
\bea
& & \de \ph_\pm = \pm i c_\pm \ph_3,\ \ \de \ph_3=0 ,\ \ \de c_\pm =0, \nn \\
& & \de {\bar c}_\pm = b_\pm, \ \ \de b_\pm =0 .
\eea
     
The Faddeev-Popov prescription consists of adding a BRST-trivial term 

\[ 
i\de (\cb_- \ph_+ + \cb_+ \ph_-)= ib_-\ph_+ 
+ ib_+\ph_- +\cb_-\ph_3\ c_+ 
-\cb_+\ph_3\ c_- 
\]
to the action in (\ref{pf}). It is now clear that the integration over $b$ will 
impose the gauge 
condition; $\ph_\pm =0$. We have  
\[
\tr \ph F= \ph_3 F_3 = \ph_3 (dA_3 + (A \wedge A)_3) = \ph_3 (dA_3 +i\stw  
A_1\wedge A_2 ) \, ,
\]
therefore, defining $\ph \equiv \ph_3, A \equiv A_3$ and $F \equiv F_3$,  the 
action in (\ref{pf}) turns into   
\[
S = \frac{1}{4\pi^2}\int_{\Sigma}\lf( \f{i}{\stw}\ph\ dA - \ph A_1\wedge A_2 
+\frac{\vep}{8} \ph^2 \ri) + \int_{\Sigma} d\mu \lf( \cb_-\ph\ c_+ 
- \cb_+\ph\ c_- \ri). 
\]
Integration over Faddeev-Popov ghosts gives

\[ 
[\det \ph^2]_{\Omega^0(\Sigma_g)},
\] 
while over $A_1$ and $ A_2$ results in 
\[ 
\lf[\det \ph^2\ri]^{-1/2}_{\Omega^1(\Sigma_g)}. 
\]

Using the Hodge decomposition theorem we can express the product of these two 
determinants as
\[ 
\f{[\det \ph^2]_{H^0(\Sigma_g)}}{[\det \ph^2]^{1/2}_{H^1(\Sigma_g)}}.
\]
When $E$ is 
a nontrivial $SO(3)$ bundle we write the curvature of the reduced $U(1)$ bundle as 
\[
 F= 2\pi (2n + 1 ) \omega + dA ,
\]
where $\omega$ is the volume form ($\int_{\Sigma} \omega =1$) and 
\[
2n + 1 =\f{1}{2\pi}\int_{\Sigma} F 
\]
is the first Chern class which characterizes the $U(1)$ bundle. 
To gauge  
fix the residual $U(1)$ symmetry
\[ 
A \to A+ d\al , 
\]
we again appeal to the Faddeev-Popov prescription. We demand that a selected 
slice  be normal to the gauge orbit,
\[
 \lan d\al, A\ran = 0,
\]
which implies that $d^\dagger A=0 $. Imposing this gauge,  the action is 
\[ 
\f{1}{4\pi^2}\int_{\Sigma}\lf(i{\stw}\pi (2n+ 1)\ph \om  +\f{\vep}{8}\ph^2
+ \f{i}{\stw}(\ph dA  + b d *A + \cb d*d c) \ri)  .
\]
The kinetic term for $A$ vanishes for $A$ a harmonic one-form, i.e.\ when $dA=0$ 
and $d^\dagger A=0$. Hence there is still a residual symmetry under  
\bea
&& A \to A + \ga \nn \\
&& b \to b + {\rm constant} \nn \\
&& c \to c + {\rm constant}, \nn
\eea
where $\ga$ is a harmonic one-form. Integration over the zero modes of $b$ and $c$ 
and over the harmonic one-forms gives an unspecified  constant factor that can 
be simply absorbed in the normalization. Therefore we need only be concerned about 
the nonzero modes. 
Dropping the harmonic part of $A$, it can be written globally and uniquely as
\[ 
A = d\al + *d\bet ,
\]
for some zero-forms $\al$ and $\bet$. The action then looks like 
\[
\f{1}{4\pi^2} \int_{\Sigma}\lf(i{\stw}\pi (2n + 1)\ph \om + \f{\vep}{8}\ph^2
 + \f{i}{\stw}(\ph d*d\bet + b d *d\al + 
\cb d*d c) \ri) , 
\]
and the measure is
\be 
\cD A =\cD \al \cD \bet\ \det[dd^\dagger ]_{\Omega_0}. \label{jac}
\ee
Note that $*^2=(-1)^p$ when acting on a $p$-form and $d^\dagger =-*d*$. 
The integral over $b$ 
and $\al$ results in a determinant, $\det [dd^\dagger ]^{-1}
_{\Omega_0}$, which cancels the jacobian in (\ref {jac}). Also the integral over 
$\bet$ gives a delta function 
\be
\de (dd^\dagger\ph)=\det [dd^\dagger]^{-1}_{\Omega_0}\ \de (\ph). \label {deter}
\ee
Notice that since we are integrating over nonzero modes the delta function 
on the right hand side is a delta function on nonconstant $\ph$'s. The 
determinant in eq. (\ref {deter}) gets cancelled against the determinant 
coming from the ghosts. At the end we are left with a finite dimensional integral 
over constant $\ph$ fields
\[ 
Z[\vep,m]= \sum_{n\in Z} \int d \ph \left( \frac{\left [\det m(1 
- \frac{2i\stw}{m}{\rm ad} \ph)\right ]
_{\Omega^1\otimes E}}{\left [\det m
(1 - \frac{2i\stw}{m}{\rm ad} \ph)\right ]_{\Omega^0\otimes E}}\right)
\f{[\det \ph^2]_
{H^0}}{[\det \ph^2]^{1/2}_{H^1}}\exp \left(-i\stw \frac{\ph (2n +1)}{4\pi}
-\frac{\vep \ph^2}{32\pi^2}\right). 
\]
Using the Riemann-Roch formula
\[
{\rm dim}\Omega^1\otimes L - {\rm dim}\Omega^0\otimes L = g-1 - c_1(L)
\] 
and the defenition of Euler characteristic of a Riemann surface
 $\ch(\Sigma_g)=2b^0-b^1
=2-2g $, and the fact that  $\ph$ is now a constant, 
we can write the partition function as 
\bea
Z[\vep,m] &=&  m^{3(g-1)} \sum_{n\in {\bf Z}} \int d\ph\ \ph^{2-2g}
(1 - \frac{8}{m^2} \ph^2)^{g-1}\lf(\f{m-2\stw \ph}{m+2\stw\ph}\ri)
^{2n + 1} \nn \\ 
&\times & \exp \left( -i\stw\frac{\ph (2n + 1 )}{4\pi}-\frac{\vep \ph^2}{32\pi^2} \right)
\eea
which is the equation (\ref{Z[m]}).

\end{document}